\documentclass[reprint,amssymb,amsmath,superscriptaddress,aps,prl]{revtex4-1}
\usepackage{graphicx}% Include figure files
\usepackage{dcolumn}% Align table columns on decimal point
\usepackage[usenames, dvipsnames]{color}
\usepackage[caption=false]{subfig}
\usepackage{bm,soul}
\usepackage[colorlinks=true,linkcolor=blue,citecolor=blue,urlcolor=blue,breaklinks=true]{hyperref}% add hypertext capabilities
\definecolor{ngreen}{rgb}{0.2,0.6,0.2}
\definecolor{amethyst}{rgb}{0.6, 0.4, 0.8}

\definecolor{golden}{rgb}{0.75,0.6,0.15}
\newcommand{\gold}{\color{golden}}

\newcommand{\ket}[1]{\ensuremath{\left| #1 \right\rangle}}
\begin{document}
\definecolor{golden}{rgb}{0.8,0.6,0.1}
\newcommand{\hmw}[1] {\textcolor{golden}{\small \bf [[#1]]}}
\newcommand{\rep}[2]{\st{#1}{\gold #2}}

%\preprint{APS/123-QED}

\title{Single-shot quantum memory advantage in the simulation of stochastic processes}
%\title{Experimental realization of a quantum epsilon machine for a 1D Ising spin chain}% Force line breaks with \\
\author{Farzad \surname{Ghafari}}
\affiliation{Centre for Quantum Dynamics, Griffith University, Brisbane, 4111, Australia}
\author{Nora Tischler}%
\email{n.tischler@griffith.edu.au }
\affiliation{Centre for Quantum Dynamics, Griffith University, Brisbane, 4111, Australia}
\author{Jayne~Thompson}%
\affiliation{Centre for Quantum Technologies, National University of Singapore, 3 Science Drive 2, Singapore, Republic of Singapore}
\author{Mile Gu}
\affiliation{Centre for Quantum Technologies, National University of Singapore, 3 Science Drive 2, Singapore, Republic of Singapore}
\affiliation{School of Physical and Mathematical Sciences, Nanyang Technological University, Singapore 639673, Republic of Singapore}
\author{Lynden K. Shalm}
\affiliation{National Institute of Standards and Technology, 325 Broadway, Boulder, Colorado 80305, USA.}
\author{Varun B. Verma}
\affiliation{National Institute of Standards and Technology, 325 Broadway, Boulder, Colorado 80305, USA.}
\author{Sae Woo Nam}
\affiliation{National Institute of Standards and Technology, 325 Broadway, Boulder, Colorado 80305, USA.}
\author{Raj B. Patel}%
\affiliation{Centre for Quantum Dynamics, Griffith University, Brisbane, 4111, Australia}
\affiliation{Clarendon Laboratory, Department of Physics, Oxford University, Parks Road OX1 3PU Oxford, United Kingdom}
\author{Howard M. Wiseman}%
%\email{e@institution.edu}
\affiliation{Centre for Quantum Dynamics, Griffith University, Brisbane, 4111, Australia}
\affiliation{Centre for Quantum Computation and Communication Technology (Australian Research Council)}
\author{Geoff J. Pryde}%
\email{g.pryde@griffith.edu.au}
\affiliation{Centre for Quantum Dynamics, Griffith University, Brisbane, 4111, Australia}

\date{\today}% It is always \today, today,

%\begin{abstract}
%\end{abstract}

%\pacs{03.67.-a,42.50.-p,03.67.Lx,03.67.Hk}% PACS, the Physics and Astronomy
 %\keywords{qqq}                           %display desired
\maketitle

%\tableofcontents

\textbf{Stochastic processes underlie a vast range of natural and social phenomena~\cite{Book-Duan2015,Book-Jacobs2010}. Some processes such as atomic decay feature intrinsic randomness, whereas other complex processes, e.g.\ traffic congestion, are effectively probabilistic because we cannot track all relevant variables. To simulate a stochastic system's future behaviour, information about its past must be stored~\cite{Crutchfield1989,Shalizi2001}and thus memory is a key resource. Quantum information processing promises a memory advantage for stochastic simulation~\cite{Gu2012,Mahoney2016,Riechers2016,Cabello2016,Thompson2017,Suen2017,Elliott2018,Aghamohammadi2018,Cabello2018,Binder2018,Thompson2018} that has been validated in recent proof-of-concept experiments~\cite{Palsson2017,Ghafari2017}. Yet, in all past works, the memory saving would only become accessible in the limit of a large number of parallel simulations~\cite{Schumacher1995,Mahoney2016}, because the memory registers of \textit{individual} quantum simulators had the same dimensionality as their classical counterparts. Here, we report the first experimental demonstration that a quantum stochastic simulator can encode the relevant information in fewer dimensions than any classical simulator, thereby achieving a quantum memory advantage even for an individual simulator. Our photonic experiment thus establishes the potential of a new, practical resource saving in the simulation of complex systems.}

\begin{figure}
	\centering
	\includegraphics[width=1\linewidth]{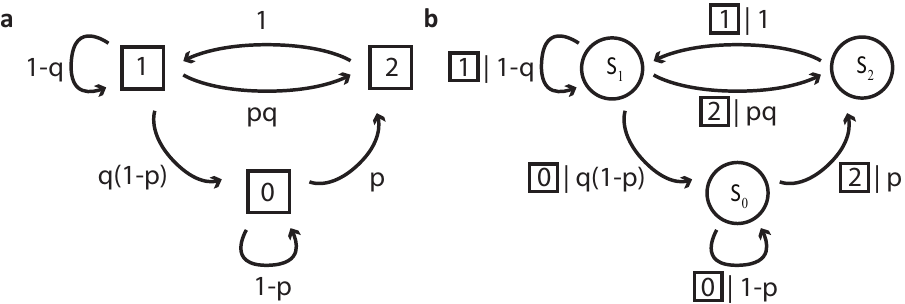}
	\caption{\textbf{The stochastic process and its simulation.} \textbf{a.} The perturbed coin process involves a coin in a box. At each step, the box is perturbed, which may or may not flip the coin. The probability of flipping from zero to one, $p$, can differ from the probability of flipping from one to zero, $q$, and similarly for the complementary probabilities of remaining in zero, $1-p$, and remaining in one, $1-q$. The process we study here is the post-processed data of the perturbed coin, which has three possible outputs at each time step, represented by the squares. The transition probabilities $ T_{ij} $, ($ i,j \in  \{0,1,2\} $), between outputs $ i $ and $ j $ are the functions of $ p $ and $ q $ provided next to the arrows. These probabilities form the transition matrix. \textbf{b.} The optimal classical simulator of the process uses causal states, as shown in the circles. The arrows represent transitions between different causal states, with the associated expressions $\boxed{j}\ |\ T_{ij}$ providing the classical output of the transition, $j$, and its probability $T_{ij}$. In this case a simple mapping exists from the past of the process to the appropriate causal state: the last output from the string of past outputs determines the causal state. The transition probability $ T_{ij} $ is the probability of transiting from causal state $ i $ to $ j $ while emitting $ j $. The eigenvalues of the transition matrix form the probability distribution of the causal states, called the stationary distribution $ \{p_i\}_{i=0,1,2} $. In the quantum case, the causal states become quantum states, $\{\ket{S_i}\}$.}
	%	The optimal classical simulator follows the same rules, except that its causal states are classical, $\{ S_i\} $, instead of quantum, $\{| S_i\rangle \}$.}
	\label{fig:process}
\end{figure}

Here we realise the first experimental demonstration of a single-shot memory advantage for simulating stochastic processes. By ``single-shot'' we mean that any individual simulator obtains an advantage, rather than requiring an asymptotically large array of simulators. We investigate a specific stochastic process, while noting that it is theoretically known that the advantage holds for a range of other simulation tasks\cite{Binder2018}. The process we simulate here can be understood as the output of a biased perturbed coin after post-processing~\cite{Thompson2018}(see Fig.\ \ref{fig:process}a): at each discrete time step, the state of the coin provides a probabilistic binary outcome, which depends on the parameters $ p $ and $ q $ that are defined by the process. Over multiple time steps, this produces a string of `zero's and `one's. Then, in post-processing, every `0' that precedes a `1' is replaced by a `2'. For classical simulation, this post-processing markedly increases the amount of past information that needs to be stored in order to generate future predictions. This is not so for quantum processors. 

It is known that for the provably optimal simulators~\cite{Thompson2018} in each class (classical or quantum) of this stochastic process (Fig.\ \ref{fig:process}b), it suffices to classify any possible past into three different states called \textit{causal states}~\cite{Shalizi2001,Gu2012}. To this end, the classical processor must have three distinguishable states, ${\left\{ {{S_i}} \right\}_{i = 0,1,2}}$, as its memory. By contrast, as we experimentally demonstrate, the quantum processor works with the three required quantum states, ${\left\{ {{|S_i\rangle}} \right\}_{i = 0,1,2}}$, compressed into a two-dimensional quantum system.

\begin{figure*}
	\centering
	\includegraphics[width=.8\linewidth]{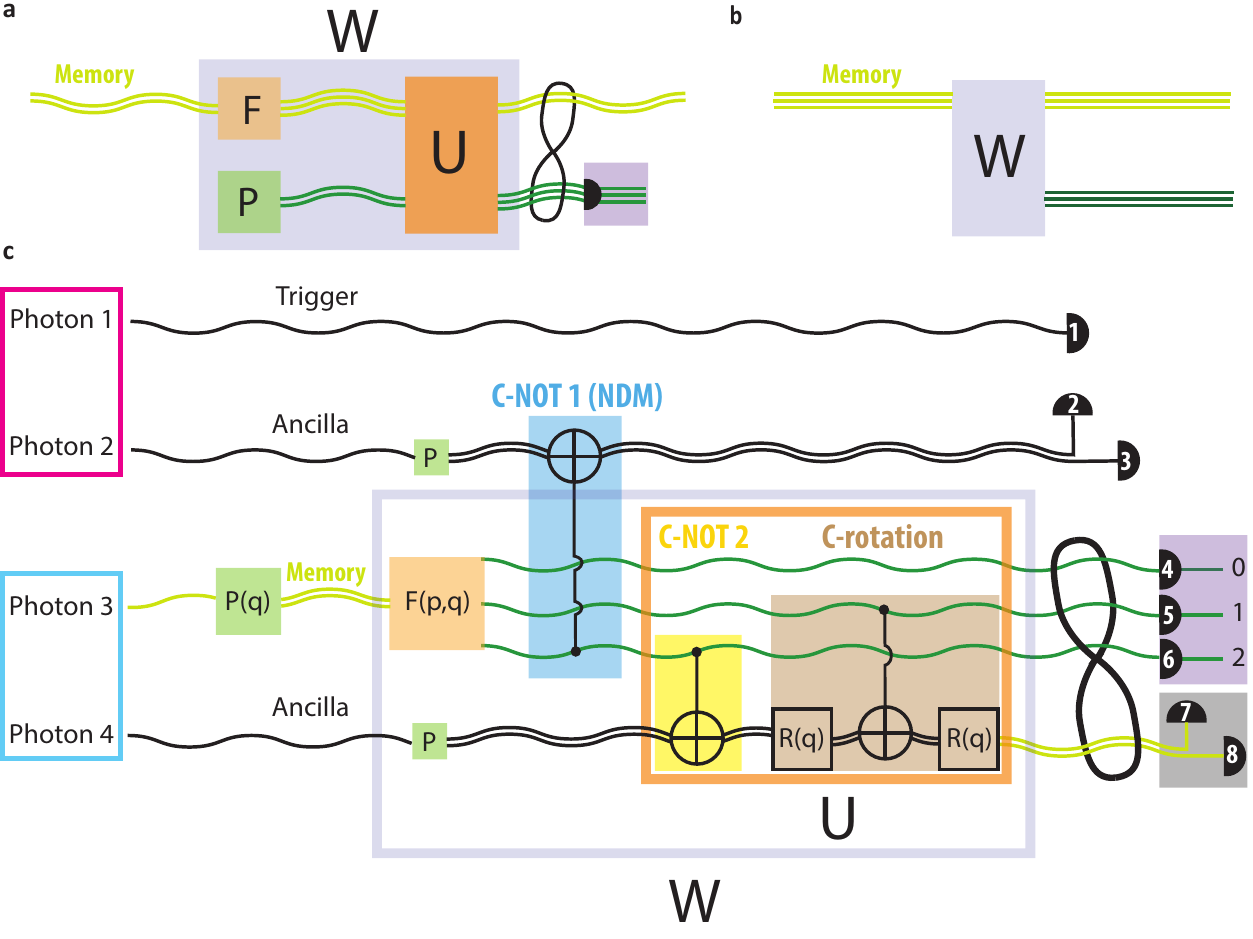}
	\caption{\textbf{Conceptual diagram of a simulation step.} \textbf{a.} The quantum simulator accepts a memory qubit and an ancilla. (We use wavy lines to denote quantum objects, with the number of lines in parallel indicating the dimensionality.) The ancilla contains no information and its preparation, $ P $, is fixed. The memory qubit undergoes a fan-out operation, $ F $, after which the information is contained in a qutrit space. Then a unitary operation, $ U $, acts on the qutrit and ancilla, outputting an entangled state of the memory qubit and a qutrit. A projective measurement of the qutrit provides the output of the simulation step and collapses the memory qubit to the appropriate state for the next step. \textbf{b.} The classical simulator requires a three-dimensional memory system. The irreversible operation $ W $ acts on the memory system to generate the classical output and the next memory state. \textbf{c.} The experimental realisation of the circuit in subfigure \textbf{a} using linear optics gates requires an ancilla qubit (photon 2) and its herald (photon 1). Following the fan-out operation $ F(p,q) $ on the memory qubit, we implement a gate, C-NOT 1, which performs a non-destructive measurement (NDM). Then the unitary operation $ U $ is performed by an additional two gates, C-NOT 2 and C-rotation. The preparation of the  memory system (photon 3) $ P(q) $, the fan-out operation $ F(p,q) $, and the single qubit rotation $ R(q) $ depend on the stochastic process parameters $ p $ and $ q $ as indicated.}
	\label{fig:conceptual}
\end{figure*}

Generally, as illustrated in Fig.\ \ref{fig:conceptual}a, a quantum simulator of a stochastic process, henceforth simply referred to as a quantum simulator, accepts a memory system and an ancilla system as inputs to a unitary transformation~\cite{Gu2012,Mahoney2016,Binder2018} for each simulation step. Of the two, only the memory system contains information about the past, while the ancilla system carries no information. The unitary transformation produces an entangled state of the output memory system and a second system. Measurement of the latter provides the output of the stochastic process, and collapses the memory system to the appropriate quantum state for the next simulation step. Importantly, the memory register enters and exits the quantum processor (W) as a two-dimensional system, unlike its classical counterpart in Fig.\ \ref{fig:conceptual}b, where the memory register is a three-dimensional system.

For the stochastic process of Fig.~\ref{fig:process}, the quantum memory required is a single qubit, in which the three causal states are encoded as three, non-mutually-orthogonal, pure quantum states, as described in Methods. We implement our simulator
% for a range of $ p $ and $ q $ values
 in a photonic quantum information processor.
%  post-selected scheme that uses four photons generated by spontaneous parametric down-conversion (SPDC).
The memory qubit is encoded in the polarisation degree of freedom of a single photon. The non-trivial unitary transformations in our experiment include a mapping from the memory qubit to a qutrit space of three spatial modes (paths), followed by a controlled-NOT (C-NOT)~\cite{O'Brien2003,Langford2005} and a controlled-rotation (C-rotation) gate, as detailed in Fig.\ \ref{fig:conceptual}c. The path measurement of this photon corresponds to measuring the qutrit in the logical basis, which provides the classical output (0, 1 or 2) of that step of the stochastic process. This collapses the output memory qubit, encoded in the polarisation state of another photon, to the correct conditional state, which can be characterised by quantum state tomography.

\begin{figure}
	\centering
	\includegraphics[width=1\linewidth]{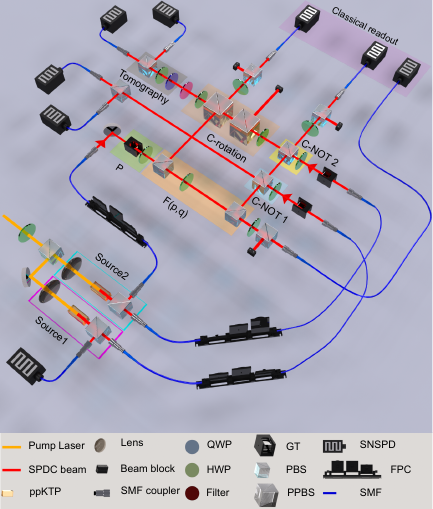}
	\caption{\textbf{Experimental setup.} Single photons are generated from SPDC events. The herald photon from Source 1 is sent straight to a heralding detector. The polarisation of the memory system is used to encode the relevant causal state in a qubit, using a half-wave plate (HWP). Ancillas are prepared in a fixed polarisation using HWPs. To implement the fanning out from the memory qubit to a qutrit, a HWP and polarising beam splitters (PBSs) are used. Each of the C-NOT 1 and C-NOT 2 gates is implemented using a HWP and a PBS. The C-rotation gate is realised via HWPs and partially polarising beam splitters (PPBSs). In order to vary the relative delay between the single photon wave packets, an automated translation stage is used to move one of the couplers. Classical readout is performed via projective measurements on the path modes of the qutrit, which collapses the memory state to the appropriate causal state. To verify the memory qubit, its state is reconstructed via quantum state tomography. A telecom bandpass filter is used in the tomography arm in order to spectrally filter the SPDC photons and maximise the visibility of the quantum interference. P stands for state preparation, SMF for single mode fibre, QWP for quarter-wave plate, GT for Glan-Taylor prism, and FPC for fibre polarisation controller. For more details, see Methods.}
	\label{fig:setup}
\end{figure}

We overcome constraints in the nondeterministic photonic implementation of consecutive quantum gates by introducing a non-destructive measurement realised by an additional C-NOT gate~\cite{Pryde2004,Ralph2006} and a corresponding ancilla photon. The photons are generated via spontaneous parametric downconversion (SPDC) and four-fold coincidences (three photons for the experiment and one ``spare'' photon to herald the presence of its pair) are detected using superconducting nanowire single-photon detectors (SNSPDs~\cite{Baek2011}) and coincidence logic modules.  The detailed experimental setup is shown in Fig\ \ref{fig:setup}, and additional details are in Methods.

The first goal of the experiment is to verify that the quantum simulator is performing the intended simulation. For this, two criteria must be fulfilled: i) After initialisation in each of the three possible causal states, the conditional output statistics, obtained through the qutrit measurement, should match the transition probabilities that determine the stochastic process (see Fig.\ \ref{fig:process}). ii) Conditioned on the qutrit measurement outcome, the correct memory state should be produced, to allow the possibility of further simulation steps.

To check the first criterion, we prepare each of the three causal states, whose definitions in terms of $ p $ and $ q $ are provided in the Methods section. For each input causal state there is a probability distribution over the three possible outputs of the stochastic process. Comparing the measured distributions with the theoretical ones, we consistently obtain (classical) fidelities~\cite{Book-Nielsen2010} above $ 0.993 $. For the second criterion, the collapsed output memory state is reconstructed by quantum state tomography, given each of the input causal states. The (quantum) fidelities of our experimental stationary states (see Methods) with the ideal stationary states are all above $0.991 $.

The second goal of the experiment is to demonstrate the quantum advantage in memory requirements. A stochastic simulator can be used in different ways, with correspondingly different ways of analysing the memory use.
%Depending on the intended use of the simulator, different types of memory can be relevant.
The most straightforward use is as a single simulator. In this scenario, the memory size, in bits, is measured by the max-entropy, which is simply $ \mathrm{log}_2D $, where $D$ is the dimensionality of the memory system~\cite{Schumacher1995,Mahoney2016}. Since the information about the past is encoded in the polarisation of a single photon, both at the beginning and at the end of the simulated step, the memory system that connects steps is obviously confined to a qubit space. In contrast to this two-level quantum system, the optimal classical simulator requires a three-level system\cite{Thompson2018}. Thus, there is a clear single-shot quantum advantage in memory.
% can be inferred.

If multiple simulations are run in parallel, the required memory is no longer determined by the dimensionality of the memory system alone. In the limit of
%an infinite
a very large number ($ N $) of parallel simulations (the independent and identically distributed (i.i.d.) case~\cite{Schumacher1995,Mahoney2016}), the minimum required memory to replicate the process faithfully is given by $ NC $, where $ C $ is called the \textit{statistical complexity}~\cite{Crutchfield1989}. The classical statistical complexity~\cite{Crutchfield1989}, $ C_{\mu} $, is the Shannon entropy of the stationary distribution over causal states, while the quantum statistical complexity~\cite{Gu2012}, $ C_Q $, is the von Neumann entropy of the quantum stationary state (see Methods for mathematical definitions).

\begin{figure*}
	\centering
	\includegraphics[width=1\linewidth]{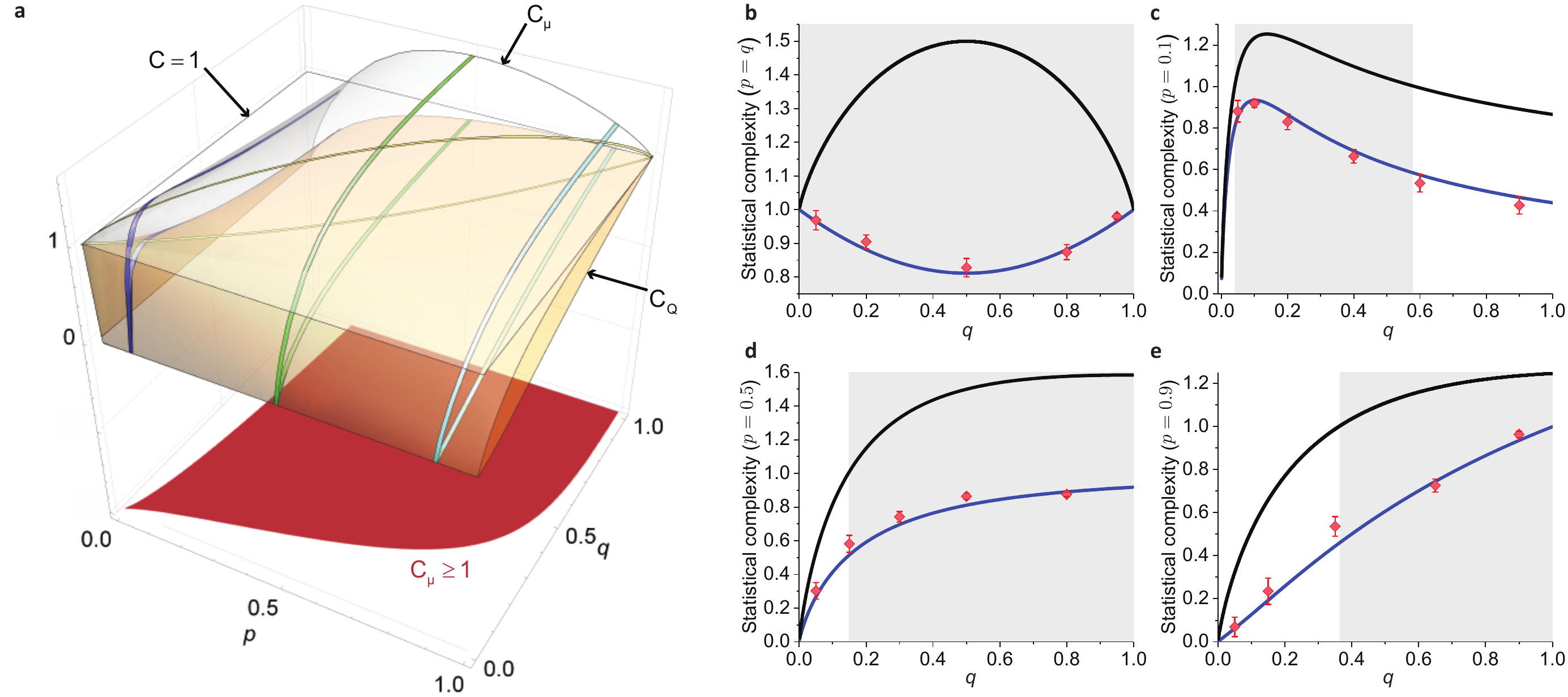}
	\caption{\textbf{Statistical complexity of the classical and quantum simulators.} \textbf{a.} The theoretically-calculated statistical complexity. The pale grey surface depicts $ C_{\mu} $, while the pale orange surface shows $ C_Q $. The transparent plane marks the value $ C=1 $. The yellow, purple, green, and cyan cuts illustrate specific cross sections, which are experimentally probed and shown in Fig.\ \ref{fig:result}b, c, d, and e, respectively. The red projection on the floor illustrates the $(p, q)$ values for which $ C_{\mu} \ge 1$. \textbf{b-e.} The quantum simulator is used to investigate several sets of processes with different values of $ p $ and $ q $. The entropy of the reconstructed stationary states (see Methods) determines the quantum statistical complexity (red dots). The black and blue curves represent the theoretical $ C_{\mu} $ and $ C_Q $, respectively. The plots demonstrate a considerable memory advantage for the i.i.d. case. Furthermore, the grey shaded areas mark processes where the complexity $ C_{\mu} $ of the classical simulator exceeds one bit, while the quantum simulation runs with only one memory qubit. % our quantum simulator assures that $ \mathrm{log}_2D=1$.
Uncertainties are estimated from the Poissonian distribution of photon counts. }
	\label{fig:result}
\end{figure*}

Fig.\ \ref{fig:result}a illustrates the theoretically-expected statistical complexities $ C_{\mu} $ and $ C_Q $ for all possible values of $  p$ and $ q $, showing the potential for a significant quantum advantage over a large region of the parameter space. We perform the simulation for sets of $ (p,q)  $ values along several cross-sections. The experimental values of $ C_Q $, shown in Fig.\  \ref{fig:result}b-e, are
%estimated based on
 determined from the density matrices of the output memory system and the transition probabilities (see Methods). The slight deviations of the experimental data compared to the theoretical curves arises from experimental imperfections such as reduced qubit purity from imperfect nonclassical interference, small imperfections and setting errors in polarisation-dependent elements,
% such as non-unit purities of the output memory states, non-ideal polarisation encoding,
and a minor imbalance in detector efficiencies. These results nevertheless demonstrate a substantial quantum advantage in the required memory for simulation in the i.i.d. case.

Thus, our quantum simulator has an advantage over its classical counterpart both for the single-shot and i.i.d.\ cases. Remarkably, we even simulate processes, marked by the shaded regions in Fig.\ \ref{fig:result}b-e, where the classical statistical complexity $C_\mu$ exceeds one bit. In these cases, we have a gap between both quantum measures and both classical measures: $C_Q < \log_2 2 < C_\mu < \log_2 3$. (Note that $\log_2 D$ always forms an upper bound on the Shannon or von Neumann entropy.)
%our \textit{single-shot} quantum simulator \blk has a lower memory requirement than the \st{\textit{average}} \blk i.i.d.\ classical simulator. This occurs wherever  Such an advantage is strictly stronger than the individual single-shot and i.i.d.\ advantages, because $ C$ is a lower bound for $ \mathrm{log}_{2}{D} $

The present experiment allows us to study both the statistical complexity and the dimensionality of the memory system. However,  for more complex processes that entail high-dimensional memory systems, the quantum state tomography required for the estimation of the statistical complexity would require increased resources (such as photons, modes, detectors), and could become prohibitively time-consuming. In contrast, verifying a dimensionality advantage remains straightforward, because it is based on counting dimensions of a Hilbert space rather than characterising quantum states. We perform a single step of the simulation in our experiment, which is already sufficient for demonstrating a quantum advantage. In the future, it would be interesting to perform multiple simulation steps with a single-shot quantum advantage.

%{\blu These results demonstrate the ability to reduce memory size in a single-shot stochastic simulation, through quantum processing.
% In instance were the causal asymmetry is large or scales with a parameter of the process, quantum models can show commensurate memory savings over the optimal classical counterparts when operating in the 'less efficient' temporal order. Indeed this experiment realizes such an instance,}
%{\blu  The dimensional advantage of quantum simulators can scale without bound. Such advantage arise naturally in the context of causal asymmetry -- a potentially unbounded memory overhead between predicting a process's future versus retrodicting its past~\cite{Crutchfield2009}. Quantum processing can drastically mitigate this gap, enabling quantum simulators that model certain processes in reverse-time with potentially unbounded reduction in dimensionality~\cite{Thompson2018}. The process presented in this paper exhibits classical causal asymmetry, and the quantum advantage in both the dimensionality and entropic costs of simulation are a direct consequence of the quantum mitigation of this effect.}

A natural question is to ask: what is the prevalence of such dimensionality advantage? While this remains an open question, its existence is certainly not isolated to the stochastic process in this experiment. Indeed, such advantage arises naturally in the context of processes that exhibit causal asymmetry---a memory overhead (in both dimensional and entropic memory costs) between predicting the future versus retrodicting the past~\cite{Crutchfield2009}. All such processes lead to dimensionality advantage, and there exist families of processes where this advantage can grow without bound~\cite{Thompson2018}.

In conclusion, we have shown that quantum information processing enables the simulation of a stochastic process with a memory that is smaller both in terms of its dimensionality (the number of orthogonal states it can support) and its von Neumann entropy, compared to the optimal classical simulator, measured by the number of states it uses and the Shannon entropy, respectively. The demonstrated decrease in the dimensionality of the memory system establishes a new type of memory saving--- namely a single-shot memory advantage. This  advantage becomes possible when the system being simulated has at least three causal states, in contrast to previous works with only two causal states~\cite{Palsson2017,Ghafari2017}. Finally, we note that although our current realisation uses nondeterministic gates, this does not affect the definition of single-shot advantage. This advantage is about the fact that multiple parallel simulators are not required, but rather that the advantage can be achieved in principle at the scale of a single simulator.

\textbf{Methods}

\textbf{Stochastic processes.}
%A stochastic process, evolving in discrete time $ t \in \mathbb{Z} $, is a collection of random variables $ X=\{...,\, X_{-1},\,X_0,X_1,\,X_2, \,...\} $, where their behaviour is determined by a joint probability $ P(X) $. At $ t=0 $, all observed variables are considered as past denoted by $ \overleftarrow{x} $, which is one of the possible configurations of $ \{...,\, X_{-1},\,X_0\} $. A faithful model is the one that take one of past events, and replicate the systems future behaviour, $ \overrightarrow{x} $, which matches the actual process's behaviour.

A stochastic process evolving in discrete time is a collection of random variables $ \{...,\, X_{t-1},\,X_t,X_{t+1},\,X_{t+2}, \,...\} $, where the previously observed variables $ \{...,\, X_{t-1},\,X_t\} $ are considered the past of the process, i.e.\ the list of past outputs. A faithful simulator is one that correctly generates the process's future statistical behaviour based on a given configuration of its past. The memory system of the simulator must store sufficient information about the past configuration to enable this faithful simulation~\cite{Shalizi2001}.
%is encoded in an input state
%according to some function
 Then, a processor
%method~\cite{Shalizi2001,Binder2018}
 acts on the memory,
generating a new classical output $X_{t+1}$ and updating the memory to be ready for the next step.

 For optimal simulation of the process that we study here~\cite{Thompson2018}, the most recent output, $ X_t $, is sufficient for determining the memory state for step $t+1$~\cite{Crutchfield1989}. The possible memory states are called causal states\cite{Crutchfield1989,Shalizi2001}, and there are three of them for this process. The classical causal states are perfectly distinguishable states, ${\left\{ {{S_i}} \right\}_{i = 0,1,2}}$.
% are given by
%\begin{equation}
%\begin{split}
%&S_0 = 0 \\
%&S_1 = 1 \\
%&S_2 = 2,
%\end{split}
%\end{equation}
%where $ 0,1 $ and $ 2 $ are elements of an orthogonal basis.
The quantum causal states, $ {\left\{ {{|S'_i\rangle}} \right\}_{i = 0,1,2}} $, can be similarly defined as
\begin{equation}
\begin{split}
&|S'_0\rangle =\sqrt{1-p} |0\rangle + \sqrt{p} |2\rangle \\
&|S'_1\rangle = \sqrt{q(1-p)} |0\rangle + \sqrt{1-q} |1\rangle + \sqrt{pq} |2\rangle \\
&|S'_2\rangle = |1\rangle,
\end{split}
\label{eq:mapstates}
\end{equation}
However, by choosing a different basis, these states
can be mapped to a single qubit space~\cite{Thompson2018}:
\begin{equation}
\begin{split}
&|S_0\rangle = |0\rangle \\
&|S_1\rangle = \sqrt{q} |0\rangle + \sqrt{1-q} |1\rangle \\
&|S_2\rangle = |1\rangle,
\end{split}
\label{eq:qcausal}
\end{equation}
where $\ket{0}, \ket{1}$ form an orthogonal basis.

\textbf{Experimental details.}

Four photons are generated via SPDC, as shown in Fig.\ \ref{fig:setup}. For this, two SPDC sources are realised using a 775 nm Ti-sapphire picosecond-pulse-length pump laser and ppKTP (46.20 $ \mu $m poling period) crystals cut for type-II collinear degenerate phase matching~\cite{Weston2016,Tischler2018}. The photons are not entangled in polarisation. The crystal temperature is controlled at $ {25^ \circ }$ C by a temperature controller. The bandpass filter is centred at 1550 nm and has a FWHM of 8.8 nm.

To run the simulator, the causal states in equation (\ref{eq:qcausal}) are encoded in the polarisation degree of freedom of a single photon acting as the memory system.  We use polarisation modes such that $ \left| 0 \right\rangle=\left| H \right\rangle  $ and $\left| 1 \right\rangle=\left| V \right\rangle $, where $ H $ and $ V $ are horizontal and vertical polarisations, respectively.

The fan-out transformation implements the basis change from equation (\ref{eq:qcausal}) to (\ref{eq:mapstates}), so that the three paths correspond to orthogonal states $\ket{0}$, $\ket{1}$, and $\ket{2}$. The experimental setup contains non-deterministic two-qubit gates. The C-NOT gates 1 and 2 are realised with a HWP, a PBS, and post-selective detection. This simplified version (compared to a universal photonic C-NOT gate~\cite{O'Brien2003}) is adequate, since the photons in the two input spatial modes always have a fixed polarisation. The controlled-rotation gate is comprised of two single-qubit rotation gates, $ R(q) $, and a two-qubit controlled-Z gate. This controlled-Z gate is based on the scheme in Ref.~\cite{Langford2005}, which uses three partially polarising beam splitters (PPBSs). However, we only require two because of the fixed polarisation in one of the input spatial modes. Four-fold coincidences are detected in a 5 ns coincidence window, using SNSPDs and fast counting electronics.
%single-photon counting modules.

  The detection channels have slightly different efficiencies, which may affect the probabilities determined from the various coincidence detection combinations and thus the inferred transition probabilities.
%  Depending on the four-fold coincidence pattern, the difference in the detection efficiencies negatively affects the statistics and thereby transition probabilities.
 The possible four-fold detection combinations are formed by coincidence detections between detectors from each of the following four sets (see Fig.\ \ref{fig:conceptual}c): $ \{1\} $, $ \{2,3\} $, $ \{4,5,6\} $ and $ \{7,8\} $. This implies that the detectors within each set should ideally have the same efficiencies. In the experiment, the detectors are installed in such a way as to match this criterion as closely as possible.
 %a set of closest detectors are chosen to be in each of mentioned the sets.

\textbf{Statistical complexity.}

The statistical complexity~\cite{Crutchfield1989,Shalizi2001,Crutchfield2009} is the minimal memory a model needs to generate future statistics correctly  using only information from past observations. The classical statistical complexity is

\begin{equation}
{C_\mu} =  - \sum_{i}^{}{{p_i}}\,\mathrm{log}_2\,{p_i},
\end{equation}
where $p_i$ is the probability of each causal state in the stationary stochastic process, i.e. in the limit of a long evolution. The quantum statistical complexity is defined as~\cite{Gu2012}:
\begin{equation}
C_Q =  - \mathrm{Tr}(\rho \log_2 (\rho)),
\label{eq:Cq}
\end{equation}
where $\rho  = \sum\limits_i {{p_i}} \left| {{S_i}} \right\rangle \left\langle {{S_i}} \right|$ is the quantum stationary state.

Our simulator implements the provably optimal model, the so-called \textit{quantum epsilon machine}~\cite{Crutchfield1989,Gu2012,Thompson2018}. Therefore, we can measure $C_Q$ by inputting the causal states described in equation (\ref{eq:qcausal}) for a given set of $ p $ and $ q $ values. The stationary state, $ \rho $, is calculated as:

\begin{equation}
\begin{split}
&\rho = {d_0}\sum\limits_{i=0}^{2} {T_{0i}\,S _{\mathrm{pol}|S_0}} \ +{d_1}\sum\limits_{i=0}^{2} {T_{1i}\,S _{\mathrm{pol}|S_1}} \, \\
&+ {d_2}\sum\limits_{i=0}^{2} {T_{2i}\,S _{\mathrm{pol}|S_2}}, \\
\end{split}
\label{eq:ensemblestate}
\end{equation}
where $ \{d_i\}_{i=0,1,2} $ are the eigenvalues of the experimentally measured transition matrix:
\begin{equation}
T=\left( {\begin{array}{*{20}{c}}
	{{T_{00}}}&{{T_{10}}}&{{T_{20}}}\\
	{{T_{01}}}&{{T_{11}}}&{{T_{21}}}\\
	{{T_{02}}}&{{T_{12}}}&{{T_{22}}},
	\end{array}} \right).
\end{equation}
$ T_{ij} $ is the probability of classical output $ j $ when the input causal state is $ |S_i\rangle $. Moreover, $ S_{\mathrm{pol|S_i}} $ is the reconstructed polarisation state of the output memory system when the input causal state is $ |S_i\rangle $.

\noindent\textbf{Data availability}

\noindent The datasets generated during and analysed in the current study are available from the corresponding author on reasonable request.

%\bibliography{3CausalState}% Produces the bibliography via BibTeX.

\noindent\textbf{Acknowledgements}

\noindent The authors would like to thank Thomas Elliott and Chengran Yang for useful discussions. This research was funded, in part, by the Australian Research Council (project no.\ DP160101911 and project no.\ CE170100012) F.G. acknowledges support by the Australian Government Research Training Program (RTP) scholarship. This work was partly supported by the National Research Foundation of Singapore and, in particular, NRF Awards  no.  NRF-NRFF2016-02, no. NRF-CRP14-2014-02, and  no. RF2017-NRF-ANR004 VanQuTe, the Singapore Ministry of Education Tier 1 RG190/17 and the John Templeton Foundation Grant no. 54914. We acknowledge the traditional owners of the land on which this work was undertaken at Griffith University, the Yuggera people.

\noindent\textbf{Author contributions}

\noindent F.G. and N.T., with contributions from R.B.P., performed the experiment, acquired the data and analysed them. L.K.S. assisted with the source design. G.J.P. supervised the experiment. N.T., F.G., G.J.P., and H.M.W. conceived the experiment, based on theoretical work by M.G. and J.T. SNSPDs were fabricated by V.B.V. and S.W.N. The manuscript was written by F.G., N.T., J.T., M.G., R.B.P., H.M.W., and G.J.P. with contributions from L.K.S., V.B.V., and S.W.N. 

\noindent\textbf{Additional information}

\noindent\textbf{Competing financial interests}

\noindent The authors declare no competing financial interests.

\end{document}